          [2001/04/25 1.1 (PWD)]
\documentclass[a4paper,twocolumn]{esapub} 
\usepackage{natbib}

\title{High-mass star formation in the Southern Hemisphere sky}
\author{V. Minier}
\affil{Service d'Astrophysique/DAPNIA/DSM/CEA CE de Saclay, 91191
Gif-sur-Yvette, France}
\author{M.G. Burton}
\author{T. Hill}
\author{C.R. Purcell}
\author{S. Longmore}
\author{A.J. Walsh}
\affil{School of Physics, University of New South Wales, Sydney
2052, Australia}
\author{F. Herpin}
\affil{L3AB, Observatoire de Bordeaux, 33270 Floirac, France}

\begin{document}

\keywords{star formation; continuum; molecular lines}

\maketitle

\begin{abstract}
We report on a multi-wavelength (IR to cm) and multi-resolution (1
mas to 20 arcsec) exploration of high-mass star formation regions in
the Galactic plane, at longitudes observable from the Southern
Hemisphere. Our source sample was originally identified through
methanol masers in the Galactic plane, which exclusively trace
high-mass star-forming regions. (Sub)millimetre continuum and
molecular line observations were carried out with SEST/SIMBA,
JCMT/SCUBA and ATNF/Mopra mm-wave telescopes and have allowed us
to identify massive ($>20$~M$_{\odot}$) and luminous ($>10^3$
L$_{\odot}$) clumps in each star-forming region. We have also
constrained the SED with additional archival IR data, the physical
conditions ($T_{dust}$, $L$, $M$) and the chemical composition of each
massive clump. Several types of objects were characterised based on
the $L_{submm}/L_{bol}$ ratio, the dust temperature and the molecular line
properties, ranging from class 0-like YSO clusters
($L_{sub}/L_{bol}\sim1\%$, $T=30$~K) to hot molecular clumps
($L_{sub}/L_{bol}\sim0.1\%$, $T=40-200$~K). Preliminary high-angular
resolution observations for a subset of the sample with the ATNF/ATCA
at 3 mm, the VLA at 15, 22 and 43 GHz and Gemini in MIR have revealed
that several (proto)stellar objects are embedded in the massive
clumps: massive protostars, hot cores and hyper-compact HII regions.
We have thus identified protoclusters of massive YSOs, which are the
precursors of the OB associations. This sample of Southern
Hemisphere star-forming regions will be extremely valuable for the
scientific preparation of the ALMA and HSO observations.
\end{abstract}

\section{Introduction}

OB stars ($>8$~M$_{\odot}$) play a major role in the chemical and
dynamical evolution of galaxies and more generally in the whole
process of star formation. However, despite being a key
astrophysical process, the overall process of high-mass star
formation is not yet well understood
. Observations have revealed
that massive stars are born in very obscured regions ($A_v>20$~mag)
of cold molecular gas and dust, far away from us (typically a few
kpc) and within dense stellar clusters ($\sim10^4$~stars~pc$^{-3}$).
They also experience short lives ($\sim10^6-10^7$ years). These
observational constraints are the main reasons of our limited
knowledge of high-mass star formation.

OB stars are also found in binaries and multiple systems within clusters. 
The multiplicity of massive stars
seems to increase with the primary mass and density of young star
clusters (Zinnecker 2003). Clarke \& Pringle (1992) suggest that the
runaway OB stars (10-25\% of O stars, 2\% of B stars) are ejected
from the clusters by binary-binary interactions and that massive
stars form in groups of binaries, which are biased toward unit mass
ratio. These elements indicate that the clustering mode is a very
likely key property of the formation and evolution of OB stars,
which might explain the difficulty to extend the standard accretion
theory of low-mass star formation to high-mass stars. Then we might
ask whether high-mass stars form in clusters from assemblies of
decoupled molecular fragments or from mergers of less massive
protostellar cores inside a large, massive infalling cloud (e.g. Padoan
et al. 2001; Bonnell \& Bate 2002).

A straightforward answer is that the standard collapse-accretion
theory with an accretion rate of $\sim10^{-5}$~M$_{\odot}$~yr$^{-1}$
cannot explain the formation of OB stars. The outward radiative
acceleration becomes important for high-mass stars as it is
proportional to the luminosity ($\propto{M_{\star}^{3.2}}$) and will
ultimately halt accretion. The infall accretion rate must then
exceed the outflow rate to produce a high-mass star, i.e. the
gravity must overcome the radiative pressure. To solve this
``accretion problem'', it has been suggested that high-mass star
formation could proceed either through protostellar mergers (Bonnell
\& Bate 2002) or through the collapse of a supersonically turbulent
core with a sufficiently large accretion rate (McKee \& Tan 2003).
In the first scenario, clusters as dense as 10$^8$~stars~pc$^{-3}$
are needed. Several starless cores will first collapse to form low-
and intermediate-mass protostars, which then fall toward the dense
inner regions of the protoclusters, to finally merge in high-mass
protostars. In the second scenario, high-mass protostars will form
after the collapse of single, massive cores and grow in mass with
accretion rates $>10^{-3}$~M$_{\odot}$~yr$^{-1}$. Yorke (2003)
suggests that the presence of an accretion disk might reduce
the effective luminosity and thus the radiative pressure on the
material in the equatorial plane. The polar component of the outward
radiative acceleration would then be greater than its equatorial
component. This would imply 2D accretion via a disk and bipolar
outflows along the disk axis.

In the past ten years, many different searches for young massive
stars have been carried out through a large range of observing methods
from radio continuum observations of compact ionised regions of gas
(ultra-compact HII regions) to spectral line (e.g. CH$_3$OH, CH$_3$CN
and NH$_3$) observations of hot molecular cores and millimetre/IR
continuum observations of embedded (proto)clusters. 
Recently, the study of methanol masers have brought new insights in
the understanding of high-mass star formation. These masers are
exclusively detected toward massive star-forming clusters (Minier et
al. 2003 and reference therein) and can potentially trace their distribution
in the Galactic Plane. Within these (proto)clusters, they
generally pinpoint hyper-compact HII regions ($<0.05$~pc in diameter),
hot molecular cores (radio-quiet and mm/MIR-bright) and sometimes
cold protostellar cores (IR-quiet and mm-bright) (e.g. Minier et al.
2004; Walsh et al. 2003). This list of objects would define an
evolutionary sequence of the early phases of high-mass
star-formation, from protostellar cores to hyper-compact HII
regions, all being precursors of HII regions.

\section{A multi-wavelength study of high-mass star formation}

An exploration of the Southern high-mass star-forming regions through methanol masers
has been undertaken in 2001 by Michael Burton's group at the University of New South
Wales, Sydney, Australia in collaboration with groups in France (P. Andr\'e), UK (M.
Thompson; M. Hoare), Chile (G. Garay) and Italy (R. Cesaroni). This work aims to make a census of 
the star-forming regions
and characterise their chemical and physical properties. Ultimately, this study will
provide the community with a large data set on high-mass star-forming regions in
the Southern sky for the Herschel Space Observatory programmes and ALMA projects on star formation.

The source sample was originally identified through methanol masers in the Galactic plane 
(Walsh et al. 1998; Minier et al. 2001), which exclusively trace high-mass star-forming regions
(Minier et al. 2003). A few ultra-compact HII (UC HII) regions without methanol masers are also 
present in the target list (Thompson et al. in prep.). In total, 131 fields at distances from 0.5
to 16 kpc were selected that can be divided in three classes: (1) fields with a methanol maser;
(2) fields with a methanol maser and a UC HII region; (3) fields with a UC HII region and without 
methanol maser. The targets are mainly located in the 1$^{st}$ and 4$^{th}$ quadrants of the Galactic
Plane.


The observations have been carried out in two phases. In {\it a first phase}, molecular
lines and continuum emission surveys were achieved to make a census of all the 
star-forming clumps and characterise their physical and chemical properties. Star-forming
clumps were identified through the emission of dust from mid-IR to millimetre wavelengths.
The SIMBA bolometer instrument on SEST at 1.2~mm ($24''$ resolution) was used to detect 
404 clumps toward the 131 fields in our source sample (Hill et al. 2005). Interestingly, 
253 secondary clumps were
imaged near those associated with the masers and/or the UC HII regions (see Fig. 1 for
an example). The dust mass was estimated and allowed us to derive the total mass ($\sim1.5\times10^3$~M$_{\odot}$)
of each clump 
(see Minier et al. 2004 for the analysis method). Hill et al. also measured and derived the
clump density ($\sim9\times10^5$~cm$^{-3}$), radius ($\sim0.5$~pc), 
visual extinction ($A_v\sim80$~mag) and surface density ($\Sigma\sim2.8$~kg~m$^{-2}$).
The average mass ($\sim9\times10^2$~M$_{\odot}$) of the secondary clumps only seen in mm is 
lower than that of the clumps associated with a maser and/or an UC HII region ($\sim2.5\times10^3$~M$_{\odot}$).
This might be due to the temperature of 20~K that was assumed for all clumps. Finally, cross-inspection
between the SIMBA source sample and the {\it MSX} catalogue in mid-IR has revealed that methanol masers and UC HII
regions are generally associated with bright mid-IR objects. However, several methanol
maser sites and mm-only sources were found to be coincident with IR dark clouds at 8 $\mu$m. 
These mm-only/IR dark clouds and maser source/IR dark clouds might represent prestellar and 
cold ``class 0'' protostellar phases of high-mass star formation.

\begin{figure}
\vspace{6.7cm}
\includegraphics{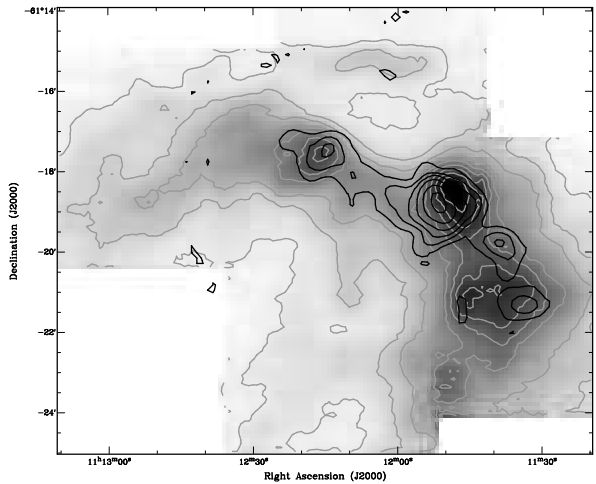}
\caption[]{NGC3576 in the 4$^{th}$ quadrant of the Galactic Plane: 
SIMBA map (contours) on the $^{13}CO$ Mopra image (grey scale). NGC3576 hosts 
two methanol maser sites and a very luminous HII region that correspond to the biggest clump. 
Secondary clumps are detected along the $^{13}CO$ filament.}
\end{figure}
 
Observations of dust emission at 450 and 850 $\mu$m (Walsh et al. 2003; Hill et al. in prep.) 
and in mid-IR ({\it MSX} catalogue) were also analysed to build the Spectral Energy Distribution (SED)
of each clump. A preliminary method has been proposed by Minier et al. (2004) who estimated the
luminosity and temperature of the clump by fitting the SED with grey body functions.
SEDs of an IR dark cloud and an IR bright clump, both associated with a methanol maser, are given in Fig. 2. 
The derived luminosities are quite similar ($\sim10^4$~L$_{\odot}$), which suggest that both clumps
host a high-mass (proto)star given the mass and the size of the clumps. However, the SEDs do not peak at
the same wavelength. The luminosity integrated over the mm/submm range represents $\sim1\%$ of the total luminosity
in the case of IR dark clouds and only $\sim0.1\%$ in the case of IR bright clumps. The $L_{submm}/L_{bol}$ ratio
for IR dark clouds is similar to what Andr\'e et al. (2000) reported for class 0 low-mass protostars. The 
IR dark cloud SED is better fitted with a single grey body function with $T_{dust}=30$~K
whilst the IR bright clump SED is fitted with two grey body functions with higher temperature. The presence
of a methanol maser implies that a protostellar object is heating up the medium to allow both radiative maser
pumping and evaporation of methanol from the icy mantles of the dust grains (Minier et al. 2003). 
Consequently, the IR dark clouds with a maser (i.e. class 0-like protostellar objects) are probably in an earlier phase of 
high-mass star formation than the IR bright clumps (i.e. hot molecular cores or/and UC HII regions).

\begin{figure*}
\vspace{7cm}
\includegraphics{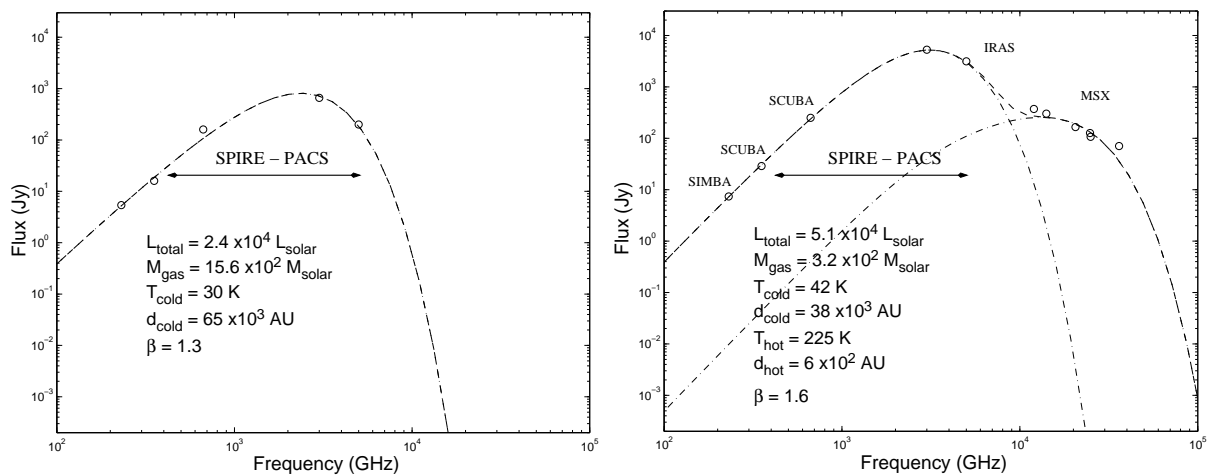}
\caption[]{Typical SED of an IR dark cloud (left) and of an IR bright clump (right). Both objects are associated
with a methanol maser. These objects emit the bulk of their luminosity between 60 and 500 $\mu$m, which will be covered
by the PACS and SPIRE instruments on Herschel with an angular resolution similar to those of SIMBA and SCUBA. At the moment,
flux densities at 60 and 100 $\mu$m are from {\it IRAS} with a few arcmin angular resolution.} 
\end{figure*}

In parallel, a molecular line survey was carried out toward a sub-sample of the SIMBA source list (Purcell
et al. in prep.). The 22m ATNF/Mopra millimetre wave telescope ($30''$ resolution) was used to identify massive 
young stellar 
objects in their hot 
molecular core phase toward 82 sources associated with Galactic methanol masers. Many of these sources are at longitudes
both observable with Northern (JCMT, IRAM) and Southern (ATCA, APEX) hemisphere telescopes. Observations of these 
sources in the 3-mm transitions of CH$_3$CN(5-4; 6-5), HCN, and HCO$^+$/H$^{13}$CO$^+$ were obtained.
Methyl cyanide emission is detected in 25 sources, of which we could derive the rotational temperatures (40-200~K)
using rotational diagram analysis. HCN and HCO$^+$ were detected in 95\% of the sources and exhibited 
signatures of infalling and expanding envelopes. 


The {\it second phase} of the observing programme is based on large scale molecular line mapping with the
ATNF/Mopra telescope and on high angular resolution observation with ATCA and IRAM-PdBI of the most 
promising objects (e.g. Fig. 1). The clumps imaged with SIMBA are identified as protoclusters in which high-mass
star formation occurs for the most massive and luminous ones (Minier et al. 2004) . The physical characteristics 
derived from single-dish telescope
observations are those of protoclusters of massive young stellar objects. Typical angular resolution
$\sim30''$ was used for the surveys described in {\it phase 1}. An order of magnitude lower ($\sim 1''$) is 
available with the current interferometers at (sub)millimetre wavelengths (ATCA-mm, IRAM-PdBI, SMA) 
and with the largest telescopes at mid-IR wavelengths (Gemini, VLT). High angular resolution observations
will reveal multiple systems of massive protostars within the clusters, and for the closest objects, these
techniques will isolate the individual protostars.

The hot molecular core and methanol maser site, G318.95-0.17 (2 kpc away), was observed 
with the newly upgraded ATCA at mm wavelengths in the 
HCO$^+$, CH$_3$OH, HCN lines. It reveals that G318.95-0.17 is a source candidate for an embedded massive 
protostar (Fig. 3). Additional observations of NH$_3$ toward the SIMBA sources are under progress.
High angular resolution observations in mid-IR and in radio continuum have also been undertaken with 
TIMMI2 and Gemini (Longmore et al. in prep.) to study the multiple systems of massive protostars inside 
each clump and derive their individual SED.  

\begin{figure*}
\vspace{12.7cm}
\includegraphics{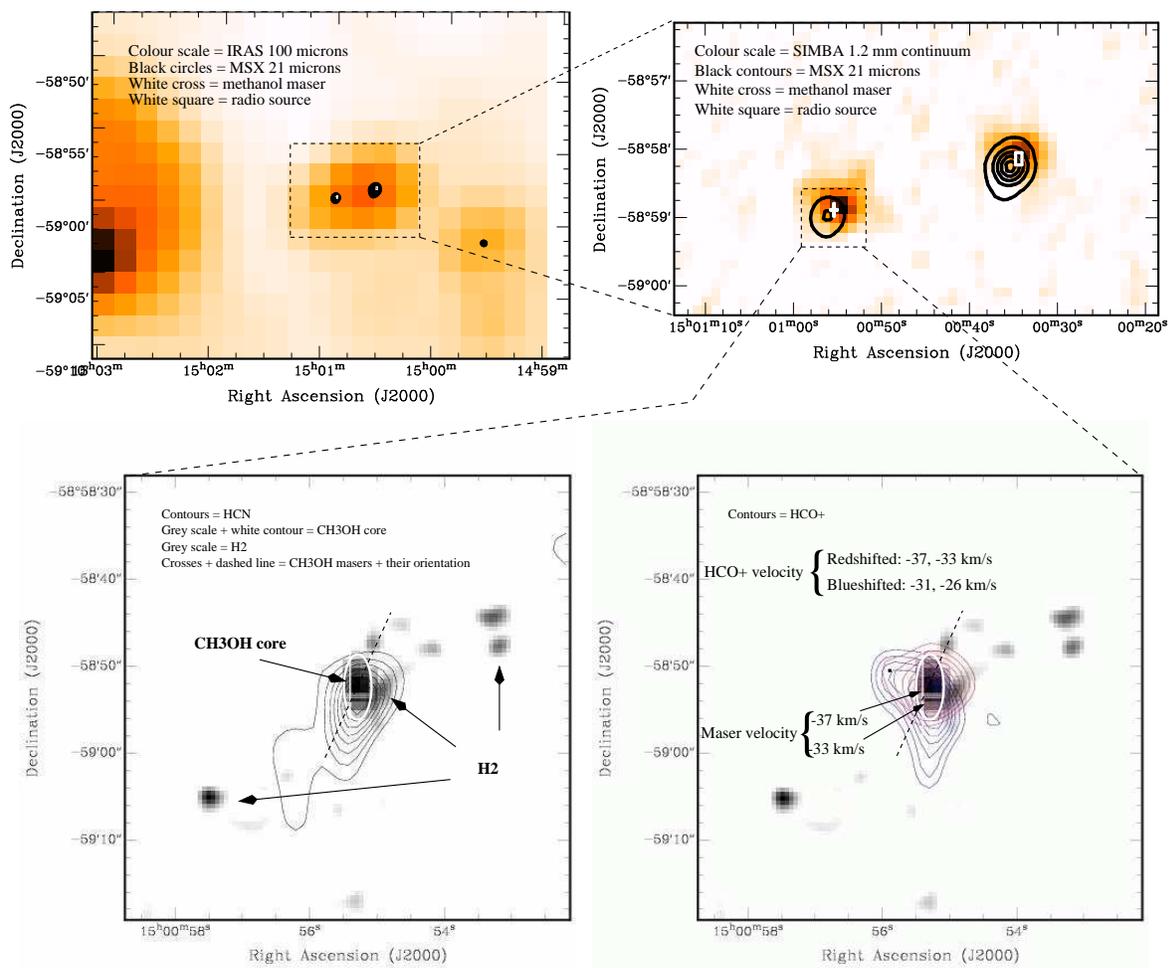}
\caption[]{Multi-wavelength and multi-resolution observations of G318.95-0.17. 
An embedded massive protostar detected nearby a strong radio, infrared source.
The site exhibits strong mid-IR, mm, methanol maser and molecular line emission, but no radio continuum. 
The HCO$^+$ emission is divided in blue- and red-shifted components, elongated in the same direction as the H$_2$
bullets observed by De Buizer (2003).}
\end{figure*}

\section*{Acknowledgments}
The author acknowledge support from the French-Australian Science and Technology (FAST) programme, 
jointly managed by the Embassy of France in Australia and the Department of Education, Science \& Training 
(DEST), Government of Australia.

\end{document}